# Numerical Analysis of the Nanoparticle Dynamics in a Viscous Liquid: Deterministic Approach

S.I. Denisov*, M.M. Moskalenko, T.V. Lyutyy, M.Yu. Baryba

*Sumy State University, 2, Rymsky-Korsakov St., 40007 Sumy, Ukraine*

We study the deterministic dynamics of single-domain ferromagnetic nanoparticles in a viscous liquid induced by the joint action of the gradient and uniform magnetic fields. It is assumed that the gradient field depends on time harmonically and the uniform field has two components, perpendicular and parallel to the gradient one. We also assume that the anisotropy magnetic field is so strong that the nanoparticle magnetization lies along the anisotropy axis, i.e., the magnetization vector is 'frozen' into the particle body. With these assumptions and neglecting inertial effects we derive the torque and force balance equations that describe the rotational and translational motions of particles. We reduce these equations to a set of two coupled equations for the magnetization angle and particle coordinate, solve them numerically in a wide range of the system parameters and analyze the role of the parallel component of the uniform magnetic field. It is shown, in particular, that nanoparticles perform only periodic rotational and translational motions if the perpendicular component of the uniform magnetic field is absent. In contrast, the nanoparticle dynamics in the presence of this component becomes non-periodic, resulting in the drift motion (directed transport) of nanoparticles. By analyzing the short and long-time dependencies of the magnetization angle and particle coordinate we show that the increase in the parallel component of the uniform magnetic field decreases both the particle displacement for a fixed time and its average drift velocity on each period of the gradient magnetic field.

## 1. INTRODUCTION

Ferromagnetic nanoparticles in viscous liquids have interesting physical properties and numerous potential and actual applications including magnetic cell separation [1, 2], magnetic hyperthermia [3, 4], and drug delivery [5, 6]. These and many other applications utilize the rotational and translational properties of nanoparticles. Some of their rotational properties induced by linearly and circularly polarized uniform magnetic fields have already been studied for different models of nanoparticles. It has been done, e.g., in Refs. [7-10] and [11-13] for the model with infinitely large and finite field of magnetic anisotropy, respectively.

One of the most common methods used to generate the translational motion of nanoparticles is the application of a gradient magnetic field [14]. The joint action of the uniform and gradient magnetic fields induces both the rotational and translational motions. It has been shown [15] that nanoparticles subjected to time-independent uniform and gradient magnetic fields can, depending on the initial particle positions, perform four regimes of their translational motion. But if nanoparticles are under the action of the uniform and time-dependent gradient magnetic fields, then their dynamics becomes much more complex [16]. The most remarkable feature that occurs in this case is the appearance of the drift motion (directed transport) of nanoparticles. It is realized in such a way that all nanoparticles with positive initial positions move to the right, and all nanoparticles with negative initial positions move to the left with different drift velocities.

In this paper, we continue the numerical study of the drift phenomenon in the deterministic approach. Our main aim is to investigate the influence of the uniform magnetic field, whose components are directed parallel and perpendicular to the gradient magnetic field, on the characteristics of the drift motion.

## 2. DETERMINISTIC EQUATIONS OF MOTION

The set of deterministic equations describing the rotational and translational motions of nanoparticles in the case of the uniform magnetic field that has only perpendicular component was recently derived in [16]. The generalization of these equations to the case when the uniform magnetic field has two components, parallel and perpendicular, is rather trivial. Therefore, here we only shortly describe the procedure of their derivation.

We assume that suspended ferromagnetic nanoparticles of radius $a$ are subjected to the harmonically oscillating gradient magnetic field

$$\mathbf{H}_g = gx \sin(\Omega t + \phi)\,\mathbf{e}_x \qquad (1)$$

and the uniform magnetic field

$$\mathbf{H} = H_\parallel \mathbf{e}_x + H_\perp \mathbf{e}_y. \qquad (2)$$

Here, $g(\geq 0)$, $\Omega$ and $\phi \in [0, \pi]$ are the gradient, frequency and initial phase of the gradient magnetic field (1), respectively, $H_\parallel$ and $H_\perp (\geq 0)$ are the parallel and perpendicular components of the uniform magnetic field (2), and $\mathbf{e}_x, \mathbf{e}_y, \mathbf{e}_z$ are the unit vectors along the corresponding axes of the Cartesian coordinate system $xyz$.

---

* denisov@sumdu.edu.ua

The magnetization $\mathbf{M} = \mathbf{M}(t)$ ($|\mathbf{M}| = M = $ const) of nanoparticles is assumed to be 'frozen' into their bodies (this approximation holds when the magnetic anisotropy field is rather large). If the initial magnetization $\mathbf{M}_0 = \mathbf{M}(0)$ lies in the $xy$ plane, then dynamics of $\mathbf{M}$ will occur in the same plane for all times, i.e.,

$$\mathbf{M} = M(\cos\varphi \, \mathbf{e}_x + \sin\varphi \, \mathbf{e}_y), \quad (3)$$

where $\varphi = \varphi(t)$ is the magnetization angle (angle between the $x$ axis and $\mathbf{M}$). In the given approximation, the magnetization dynamics of each nanoparticle is described by the kinematic equation

$$\frac{d}{dt}\mathbf{M} = \boldsymbol{\omega} \times \mathbf{M}. \quad (4)$$

Here, $\boldsymbol{\omega} = \boldsymbol{\omega}(t)$ is the particle angular velocity and the sign $\times$ denotes the vector product. As it follows from (3) and (4), $\boldsymbol{\omega} = \omega_z \mathbf{e}_z$ (the nanoparticle rotates about the $z$ axis) and $\omega_z = d\varphi/dt$.

Neglecting the inertial effects, the rotational and translational motions of nanoparticles are described by a set of torque $\mathbf{t}_d + \mathbf{t}_f = 0$ and force $\mathbf{f}_d + \mathbf{f}_f = 0$ balance equations, where indexes $d$ and $f$ correspond to driving and frictional values, respectively. In dilute suspensions, when the interparticle interactions are small, for the driving torque and force in our case we have $\mathbf{t}_d = V\mathbf{M} \times (\mathbf{H}_g + \mathbf{H})|_{x=R_x}$ and $\mathbf{f}_d = V(M_x \partial/\partial x)\mathbf{H}_g|_{x=R_x}$. Here, $V = (4/3)\pi a^3$ is the nanoparticle volume and $R_x = R_x(t)$ denotes the $x$-coordinate of the particle center. Expressions (1) and (2) show that the driving torque is given by $\mathbf{t}_d = t_d \mathbf{e}_z$ with

$$t_d = MV[H_\perp \cos\varphi - H_\parallel \sin\varphi - gR_x \sin\varphi \sin(\Omega t + \phi)]. \quad (5)$$

Similarly, from (1) and $M_x = M\cos\varphi$ it follows that $\mathbf{f}_d = f_d \mathbf{e}_x$ with

$$f_d = MVg \cos\varphi \sin(\Omega t + \phi). \quad (6)$$

If the Reynolds rotational and translational numbers are small, then the frictional torque and force are expressed as $\mathbf{t}_f = -6\eta V\boldsymbol{\omega}$ and $\mathbf{f}_f = -6\eta\pi a\mathbf{V}$ [17]. Here, $\eta$ designates the dynamic viscosity of the liquid and $\mathbf{V} = (dR_x/dt)\mathbf{e}_x$ is the translational particle velocity. Using the above results and the relation $\boldsymbol{\omega} = (d\varphi/dt)\mathbf{e}_z$, the torque balance equation $\mathbf{t}_d + \mathbf{t}_f = 0$ reduces to

$$\frac{d\varphi}{dt} = \frac{MV}{6\eta}[H_\perp \cos\varphi - H_\parallel \sin\varphi - gR_x \sin\varphi \sin(\Omega t + \phi)]. \quad (7)$$

At the same time, the force balance equation $\mathbf{f}_d + \mathbf{f}_f = 0$ yields

$$\frac{dR_x}{dt} = \frac{2Mga^2}{9\eta}\cos\varphi \sin(\Omega t + \phi). \quad (8)$$

Introducing the dimensionless time $\tau = \Omega t$, particle coordinate $r_x = r_x(\tau) = R_x/a$ and frequencies

$$\nu_g = \frac{Mga}{6\eta\Omega}, \quad \nu_\perp = \frac{MH_\perp}{6\eta\Omega}, \quad \nu_\parallel = \frac{MH_\parallel}{6\eta\Omega}, \quad (9)$$

Eqs. (7) and (8) are reduced to a set of dimensionless equations with respect to $\varphi$ and $r_x$

$$\dot{\varphi} = \nu_\perp \cos\varphi - \nu_\parallel \sin\varphi - \nu_g r_x \sin\varphi \sin(\tau + \phi), \quad (10)$$

$$\dot{r}_x = (4/3)\nu_g \cos\varphi \sin(\tau + \phi), \quad (11)$$

These equations describe the coupled rotational and translational motions of particles with the initial conditions $\varphi_0 = \varphi(0) \in [0, \pi]$ and $r_{x0} = r_x(0) \in (-\infty, \infty)$. It should be noted that, according to Eqs. (10) and (11), the gradient magnetic field $\mathbf{H}_g$ influences the translational motion of nanoparticles directly, while the uniform magnetic field $\mathbf{H}$ influences indirectly.

Equations (10) and (11) allow us to find the connection between their solution $\{\varphi, r_x\}_{\nu_\parallel}$ for a fixed $\nu_\parallel$ and their solution $\{\tilde{\varphi}, \tilde{r}_x\}_{\tilde{\nu}_\parallel}$ for a fixed $\tilde{\nu}_\parallel = -\nu_\parallel$. We assume that these solutions satisfy the initial conditions $\{\varphi_0, r_{x0}\}_{\nu_\parallel}$ and $\{\tilde{\varphi}_0, \tilde{r}_{x0}\}_{\tilde{\nu}_\parallel}$, respectively. Then, choosing $\tilde{\varphi} = \pi - \varphi$ with $\tilde{\varphi}_0 = \pi - \varphi_0$ and $\tilde{r}_x = -r_x$ with $\tilde{r}_{x0} = -r_{x0}$, it is not difficult to verify that $\{\tilde{\varphi}_0, \tilde{r}_{x0}\}_{\tilde{\nu}_\parallel = -\nu_\parallel}$ is indeed the solution of Eqs. (10) and (11). Therefore, without loss of generality, we may restrict our further analysis to $\nu_\parallel \geq 0$.

## 3. NUMERICAL RESULTS

Taking into account the parameters of ferromagnetic materials [18] and possible parameters of the applied magnetic fields, we numerically solved Eqs. (10) and (11) for a wide range of the model parameters: $\nu_g \lesssim 1$, $\nu_\perp \in (0, 10^2)$, $\nu_\parallel \in (0, 10^2)$ and $\tau \in (0, 10^6)$.

### 3.1 The Case with $\nu_\perp = 0$ and $\nu_\parallel = 0$

According to [19], there is no drift motion of nanoparticles at $\nu_\perp = 0$ and $\nu_\parallel = 0$. In this case, both the particle coordinate $r_x$ and the magnetization angle $\varphi$ are periodic functions of time $\tau$. For illustration, in Fig. 1 we show these functions for a given set of the model parameters.

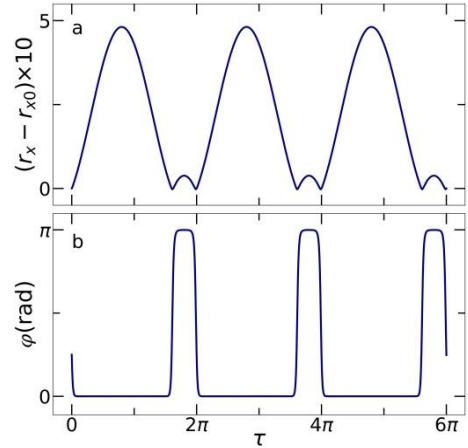

**Fig. 1** – The time dependencies of the particle coordinate $r_x$ (a) and the magnetization angle $\varphi$ (b) for $\nu_g = 0.2$, $\nu_\perp = 0$, $\nu_\parallel = 0$, $r_{x0} = 2 \times 10^2$, $\varphi_0 = \pi/4$ and $\phi = \pi/5$

### 3.2 The Case with $\nu_\perp = 0$ and $\nu_\parallel > 0$

Our numerical analysis shows that in this case the drift motion of nanoparticles is also absent. However, the parallel component of the uniform magnetic field changes qualitatively the nanoparticle dynamics. The most significant change occurring at $\nu_\parallel \neq 0$ is that the solution of Eqs. (10) and (11) tends with time to the steady state one $\{\varphi_{st}, r_{xst}\}$, where depending on the model parameters the functions $\varphi_{st}$ and $r_{xst}$ are given by

$$\varphi_{st} = 0, \quad r_{xst} = (4/3)v_g[\cos\phi - \cos(\tau + \phi)] \quad (12)$$

or

$$\varphi_{st} = \pi, \quad r_{xst} = -(4/3)v_g[\cos\phi - \cos(\tau + \phi)]. \quad (13)$$

As seen, in this state all particles perform the same harmonic oscillations (they do not depend on $r_{x0}$). The steady state solutions (12) and (13) are established during some transition time $\tau_{tr}$ that strongly depends on $v_\parallel$. In particular, if the value of $v_\parallel$ is very small, then the steady state solution is established during many periods of the gradient magnetic field. In contrast, $\tau_{tr} \ll 1$ if $v_\parallel$ is rather large. To illustrate this fact, in Fig. 2a we show the time dependence of the particle coordinate $r_x$ obtained numerically (solid line) and the theoretical result for $r_{xst}$ from (12) (dotted line). Here, $v_\parallel = 0.6$ and the other parameters are chosen to be the same as in Fig. 1. The time dependence of the magnetization angle $\varphi$ is shown in Fig. 2b for $v_\parallel = 0.6$ and $v_\parallel = 3$. From the results presented in Fig. 2 it follows that $\tau_{tr} \sim 6\pi$ for $v_\parallel = 0.6$ and $\tau_{tr} \ll 1$ for $v_\parallel = 3$.

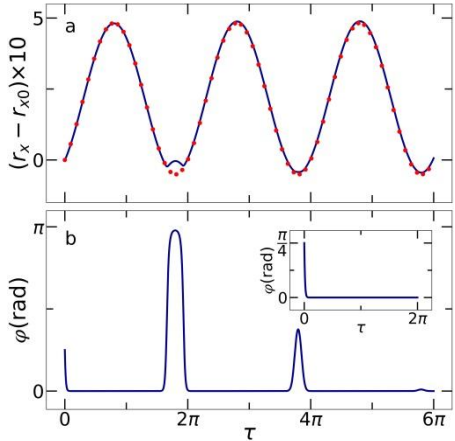

**Fig. 2** – The time dependencies of the particle coordinate $r_x$ (a) and the magnetization angle $\varphi$ (b) for $v_\parallel = 0.6$. The other parameters are the same as in Fig. 1. The dotted line corresponds to the theoretical result (12) for $r_{xst}$. The inset in Fig. 2b: the dependence of $\varphi$ on time $\tau$ at $v_\parallel = 3$

### 3.3 The Case with $v_\perp > 0$ and $v_\parallel \geq 0$

It is known [16] that at $v_\perp > 0$ and $v_\parallel = 0$ the drift motion of nanoparticles appears. This phenomenon occurs due to the coupling between their rotational and translational motions. Since as mentioned above the parallel component of the uniform magnetic field strongly influences the rotational properties of nanoparticles (if they are not too far from the origin), one may expect that it affects the drift motion as well.

We start our analysis with studying the dependence of the nanoparticle displacement $r_x - r_{x0}$ on $v_\perp$ during a short time $\tau = 2\pi N$ ($N$ is a small natural number) at $v_\parallel = 0$. In Fig. 3, this dependence is shown for $N = 8$, the other parameters are given in the figure caption. The most striking feature of this dependence is that it has the maximum value $\max[r_x(2\pi N) - r_{x0}]$ at $v_\perp = v_{\perp\max}$ (in the considered case $\max[r_x(16\pi) - r_{x0}] = 3.4$ and $v_{\perp\max} = 1.4$). Note also that with increasing $v_\perp$ the displacement $r_x(16\pi) - r_{x0}$ initially sharply increases and then, after reaching the peak value, slowly decreases (e.g., $r_x(16\pi) - r_{x0} = 3.4 \times 10^{-1}$ at $v_\perp = 50$).

Next, we studied the dependencies of the particle coordinate $r_x$ and the magnetization angle $\varphi$ on time $\tau$ (at short time intervals) for nanoparticles near the origin. To illustrate the obtained results, in Fig. 4 we show these dependencies for $v_\perp = 10$ and different values of the parameter $v_\parallel$. These results demonstrate a common tendency: an increase in the parallel component of the uniform magnetic field (i.e., the increase of $v_\parallel$) decreases the growth of $r_x$. To characterize this decrease quantitatively, we introduce the dimensionless average drift velocity of nanoparticles $\bar{v}_n = [r_x(2\pi n) - r_x(2\pi n - 2\pi)]/2\pi$ ($n = 1, 2, ..., N$) on the $n$-th period of the gradient magnetic field. Since the condition $\bar{v}_n = $ const holds if $N$ is rather small (see also below), we can characterize the average drift velocity on the time interval $(0, 2\pi N)$ by the formula $\bar{v} = [r_x(2\pi N) - r_{x0}]/2\pi N$. Thus, an increase in the parameter $v_\parallel$ decreases the drift velocity of nanoparticles. For data from Fig. 4 we have $\bar{v} = 3.1 \times 10^{-2}$ for $v_\parallel = 0$, $\bar{v} = 2.4 \times 10^{-2}$ for $v_\parallel = 5$ and $\bar{v} = 1.3 \times 10^{-2}$ for $v_\parallel = 10$.

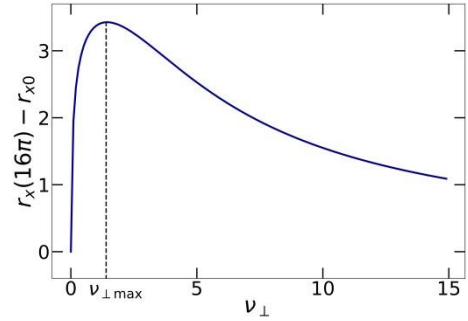

**Fig. 3** – The dependence of the nanoparticle displacement $r_x - r_{x0}$ during time $\tau = 2\pi N$ on $v_\perp$. The model parameters are chosen as follows: $N = 8$, $v_g = 0.1$, $v_\parallel = 0$, $r_{x0} = 50$, $\varphi_0 = \pi/4$ and $\phi = \pi/5$

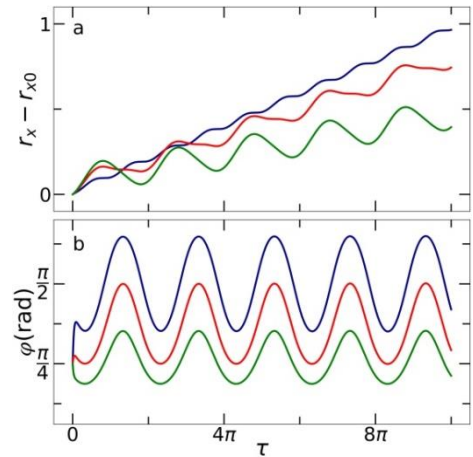

**Fig. 4** – The time dependencies of the particle coordinate $r_x$ (a) and the magnetization angle $\varphi$ (b) for the following model parameters: $N = 5$, $v_g = 0.1$, $v_\perp = 10$, $r_{x0} = 50$, $\varphi_0 = \pi/4$, $\phi = \pi/5$ and $v_\parallel = 0$ (blue line), $v_\parallel = 5$ (red line), $v_\parallel = 10$ (green line)

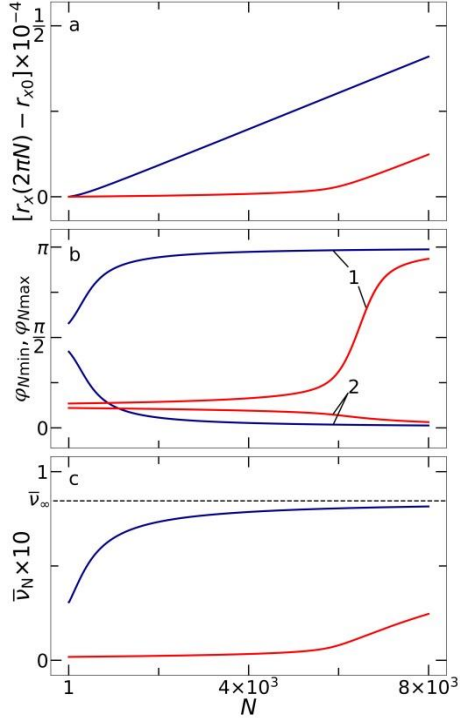

**Fig. 5** – The particle coordinate $r_x(2\pi N)$ after $N$ periods of the gradient magnetic field (a), the maximal $\varphi_{N\max}$ (curves 1) and minimal $\varphi_{N\min}$ (curves 2) values of the magnetization angle on the $N$-th period (b) and the average drift velocity $\bar{v}_N$ of the particle on the $N$-th period (c). These numerical results are obtained for $v_g = 0.1$, $v_\perp = 10$, $r_{x0} = 50$, $\varphi_0 = \pi/4$, $\phi = \pi/5$ and $v_\parallel = 0$ (blue lines), $v_\parallel = 25$ (red lines)

In general, for a given particle its drift velocity depends not only on the model parameters, but also on time during which the particle moves. This occurs because the particle shifts to the right or left (the direction depends on the sign of $r_{x0}$) during each period of the gradient magnetic field. As a consequence, the gradient field acting on the particle increases and the rotational and translational properties of this particle, including its drift velocity, change with time. However, if $r_{x0}$ or $r_x(2\pi N)$ (the number of periods $N$ can be very large if the initial particle position is not too far from the origin) are so large that the gradient magnetic field essentially exceeds the uniform one, then all such particles move with the same time independent drift velocity [16]

$$\bar{v}_\infty = \text{sgn}(r_{x0}) \frac{8 v_g}{3\pi} \qquad (14)$$

(sgn denotes the sign of $r_{x0}$), which does not depend on the uniform magnetic field. In contrast, according to the above results, this field, including its parallel component, strongly affects the particle coordinate $r_x(2\pi N)$.

This preliminary analysis is confirmed by the numerical results presented in Fig. 5. The fact that the parameter $v_\parallel$ decreases the coordinate $r_x(2\pi N)$ of a given particle is illustrated in Fig. 5a, where the dependencies of $r_x(2\pi N)$ on $N$ are shown for $v_\parallel = 0$ (blue line) and $v_\parallel = 25$ (red line). The plots of the maximal and minimal values of the magnetization angle on the $N$-th period of the gradient magnetic field, $\varphi_{N\max}$ and $\varphi_{N\min}$, are depicted in Fig. 5b. Here, $\varphi_{1\max} = 2.0$,

$\varphi_{1\min} = 1.1$ for $v_\parallel = 0$, and $\varphi_{1\max} = 4.6 \times 10^{-1}$, $\varphi_{1\min} = 3.2 \times 10^{-1}$ for $v_\parallel = 25$. In the long-time limit (when $N \to \infty$), $\varphi_{N\max} \to \pi$ and $\varphi_{N\min} \to 0$ for any fixed uniform magnetic field. Finally, in Fig. 5c we show the average drift velocity $\bar{v}_N$ of a given particle on the $N$-th period of the gradient magnetic field. For the used parameters we have $\bar{v}_1 = 3.1 \times 10^{-2}$ at $v_\parallel = 0$ and $\bar{v}_1 = 1.7 \times 10^{-3}$ at $v_\parallel = 25$. As seen, if $N$ is not too large, then $v_\parallel$ strongly decreases the drift velocity, but in the long-time limit the drift velocity does not depend on $v_\parallel$ and approaches $\bar{v}_\infty = 8.5 \times 10^{-2}$. We note, however, that the more $v_\parallel$ is, the slower $\bar{v}_N$ grows with increasing $N$.

## 4. CONCLUSIONS

We have numerically investigated the dynamics of single-domain ferromagnetic nanoparticles induced by the harmonically oscillating gradient magnetic field and uniform magnetic field with two components, perpendicular and parallel to the gradient one. The set of the coupled first-order differential equations, describing the rotational and translational motions of such particles, has been generalized to the case, when the parallel component of the uniform magnetic field is nonzero. Solving this set of equations numerically, we have shown that, in contrast to the joint action of the perpendicular and gradient fields, the joint action of the parallel and gradient fields does not induce the drift motion of particles. In the last case, the particle coordinate and the magnetization angle tend with time to the steady state ones. We have established that the transition time, which is necessary to reach these states, strongly decreases with increasing the parallel field value. In the long-time limit, all particles perform the same harmonic oscillations.

If the perpendicular component of the uniform magnetic field is nonzero, then the drift and periodic motions of nanoparticles exist simultaneously. In contrast to the previous case, the intervals of the particle oscillations are not localized near their initial positions, vice versa they are displaced to the right or left of the origin. This means that even if the initial particle positions are not too far from the origin, the role of the gradient magnetic field grows with time (due to the drift phenomenon) and becomes dominant at long times. However, the uniform magnetic field, including its parallel component, strongly influences the time dependence of the average particle coordinate. In contrast, the average drift velocity of a given particle greatly decreases with increasing the value of the parallel component only if the drift time is not too large. In the long-time limit, when the role of the uniform magnetic field is negligible, the drift velocity approaches the limiting one.

**ACKNOWLEDGEMENTS**

This work was partially supported by the Ministry of Education and Science of Ukraine under Grant No. 0119U100772. The authors also acknowledge the support of the Ukrainian State Fund for Fundamental Research under Grant No. F 81/41894.

**REFERENCES**

1. *Magnetic Cell Separation* (Eds. by M. Zborowski, J.J. Chalmers) (Amsterdam: Elsevier: 2008).
2. B.D. Plouffe, S.K. Murthy, L.H. Lewis, *Rep. Prog. Phys.* **78**, 016601 (2015).
3. P. Das, M. Colombo, D. Prosperi, *Colloids Surf. B* **174**, 42 (2019).
4. X. Liu at all, *Theranostics* **10**, 3793 (2020).
5. *Nanoparticles for Biomedical Applications: Fundamental Concepts, Biological Interactions and Clinical Applications* (Eds. by E.J. Chung, L. Leon, C. Rinaldi) (Amsterdam: Elsevier: 2020).
6. *Nanoparticles and their Biomedical Applications* (Ed. by A.K. Shukla) (Singapore: Springer: 2020).
7. G. Bertotti, I.D. Mayergoyz, C. Serpico, *Nonlinear Magnetization Dynamics in Nanosystems* (London: Elsevier: 2009).
8. T.V. Lyutyy, S.I. Denisov, V.V. Reva, Yu.S. Bystrik, *Phys. Rev. E* **92**, 042312 (2015).
9. K.D. Usadel, *Phys. Rev. B* **95**, 104430 (2017).
10. T.V. Lyutyy, V.V. Reva, *Phys. Rev. E* **97**, 052611 (2018).
11. N.A. Usov, B.Ya. Liubimov, *J. Appl. Phys.* **112**, 023901 (2012).
12. K.D. Usadel, C. Usadel, *J. Appl. Phys.* **118**, 234303 (2015).
13. T.V. Lyutyy, O.M. Hryshko, M.Yu. Yakovenko, *J. Magn. Magn. Mater.* **473**, 198 (2019).
14. J. Svoboda, *Magnetic Techniques for the Treatment of Materials* (Dordrecht: Kluwer: 2004).
15. S.I. Denisov, T.V. Lyutyy, M.O. Pavlyuk, *J. Phys. D: Appl. Phys.* **53** 405001 (2020).
16. S.I. Denisov, T.V. Lyutyy, A.T. Liutyi, *J. Phys. D: Appl. Phys.* **55** 045001 (2022).
17. S.I. Rubinow, J.B. Keller, *J. Fluid Mech.* **11**, 447 (1961).
18. A.P. Guimarães, *Principles of Nanomagnetism, 2ⁿᵈ Edition* (Cham: Springer: 2017).
19. S.I. Denisov, T.V. Lyutyy, A.T. Liutyi, *J. Nano- Electron. Phys.* **12**, 06028 (2020).

## Чисельний аналіз динаміки наночастинок у в'язкій рідині: Детерміністичний підхід

С.І. Денисов, М.М. Москаленко, Т.В. Лютий, М.Ю. Бариба

*Сумський державний університет, вул. Римського-Корсакова, 2, 40007 Суми, Україна*

Чисельно вивчається механічна динаміка однодоменних феромагнітних наночастинок у в'язкій рідині, яка індукується сумісною дією градієнтного магнітного поля, що змінюється у часі за гармонічним законом, та однорідного магнітного поля, що має перпендикулярну та паралельну до напрямку градієнтного поля компоненти. Використовуючи наближення невзаємодіючих жорстких диполів, у відповідності з яким вектор намагніченості частинки вважається 'вмороженим' в її тіло, та нехтуючи інерційними ефектами, узагальнено систему двох диференційних рівнянь для кута намагніченості та координати наночастинки, що описують її обертальний та трансляційний рухи у цьому випадку. Отриману систему рівнянь розв'язано чисельно для широкого кола параметрів задачі та проведено порівняльний аналіз динаміки наночастинок в залежності від величини паралельної компоненти однорідного магнітного поля. Встановлено, зокрема, що обертальний та трансляційний рухи наночастинок є суто періодичними, якщо перпендикулярна компонента однорідного магнітного поля відсутня. Якщо ж ця компонента присутня, тоді динаміка наночастинок стає неперіодичною і з'являється їх дрейфовий рух (спрямований транспорт). Шляхом аналізу часових залежностей кута намагніченості та координати наночастинок на коротко та довгострокових часових інтервалах встановлено, що зростання величини паралельної компоненти суттєво зменшує як переміщення частинок за фіксований час, так і їх середню швидкість на кожному періоді градієнтного магнітного поля.

**Ключові слова:** Однодоменні наночастинки, В'язка рідина, Однорідне та градієнтне магнітні поля, Трансляційне та обертальне рівняння руху, Дрейф наночастинок.